\documentclass[a4paper,11pt]{article}
\pdfoutput=1
\usepackage{jcappub}
\usepackage{amssymb}
\usepackage{amsmath}
\usepackage{mathrsfs}
\usepackage{bm}
\usepackage{graphicx}
\usepackage{hyperref}

\newcommand{\erfc}{\mathrm{erfc}}
\newcommand{\Mpc}{\mathrm{Mpc}}

\begin{document}

\title{Primordial black holes and gravitational waves from parametric amplification of curvature perturbations}

\author[a,b,c]{Rong-Gen Cai,}
\author[a,b,c]{Zong-Kuan Guo,}
\author[a,b]{Jing Liu,}
\author[a,b]{Lang Liu,}
\author[a,b,1]{Xing-Yu Yang\note{Corresponding author.}}

\affiliation[a]{
	CAS Key Laboratory of Theoretical Physics, Institute of Theoretical Physics, Chinese Academy of Sciences, P.O. Box 2735, Beijing 100190, China
}
\affiliation[b]{
	School of Physical Sciences, University of Chinese Academy of Sciences, No.19A Yuquan Road, Beijing 100049, China
}
\affiliation[c]{
    School of Fundamental Physics and Mathematical Sciences, Hangzhou Institute for Advanced Study, University of Chinese Academy of Sciences, Hangzhou 310024, China
}

\emailAdd{cairg@itp.ac.cn}
\emailAdd{guozk@itp.ac.cn}
\emailAdd{liujing@itp.ac.cn}
\emailAdd{liulang@itp.ac.cn}
\emailAdd{yangxingyu@itp.ac.cn}

\abstract{
    We investigate a new mechanism to create large curvature perturbations on small scales due to parameter resonance in a single-field inflationary model with a small periodic structure upon the potential.
    After reentering the horizon, the amplified curvature perturbations can lead to observable primordial black holes as well as stochastic gravitational waves.
    The mass of primordial black holes and frequency of the induced gravitational waves depend on the model parameters.
    The resulted primordial black hole could constitute all dark matter or a fraction of dark matter in the universe, and corresponding stochastic gravitational waves fall in the frequency band measurable for the pulsar timing array and the space-based gravitational wave detectors.
}

\maketitle

\section{Introduction}
The overdense regions in the early universe could gravitationally collapse to form black holes~\cite{Zeldovich:1967,Hawking:1971ei,Carr:1974nx}.
Such black holes are usually called primordial black holes (PBHs) and they offer us an opportunity to explore the physics of the early universe.
Recently, PBHs have attracted a lot of attention and have been studied extensively~\cite{Khlopov:2008qy,Sasaki:2018dmp,Germani:2018jgr,Bartolo:2018evs,Bartolo:2018rku,Mishra:2019pzq}, since they can be the candidates for dark matter (DM)~\cite{Green:2004wb,Frampton:2009nx,Carr:2009jm,Carr:2016drx,Gao:2018pvq,Nakama:2019htb,Fu:2019ttf}, the seeds for galaxy formation~\cite{Bean:2002kx,Kawasaki:2012kn,Nakama:2017xvq,Carr:2018rid}, and even the sources of LIGO/VIRGO detection~\cite{Bird:2016dcv,Sasaki:2016jop}, depending on their mass and abundance, which can be constrained by a plenty of observations.

It is well known that the scalar perturbations can be the source of the tensor perturbations at the nonlinear order.
If the scalar perturbations on small scales during inflation are large enough, they would lead to the formation of PBHs, when they reenter the horizon.
On the other hand, those large curvature perturbations, as source of the tensor perturbations, would result in an isotropic stochastic gravitational wave (GW) background observable today~\cite{Ananda:2006af,Baumann:2007zm,Garcia-Bellido:2016dkw,Inomata:2016rbd,Garcia-Bellido:2017aan,Kohri:2018awv,Cai:2018dig,Inomata:2018epa,Cai:2019amo,Inomata:2019zqy,Inomata:2019ivs,Yuan:2019udt,Cai:2019elf,Cai:2019cdl,Yuan:2019wwo,Chen:2019xse,DeLuca:2019ufz,Inomata:2019yww,Bhattacharya:2019bvk,Domenech:2019quo,Fu:2019vqc}.

Primordial curvature perturbations from quantum fluctuations during inflation can seed large-scale structure and explain the cosmic microwave background(CMB) temperature anisotropy successfully~\cite{Lewis:1999bs,Bernardeau:2001qr}.
On CMB scales the amplitude of the power spectrum of curvature perturbations is found to be around $2.2\times10^{-9}$, however, on relatively small scales the constraints on the curvature perturbations are very loose~\cite{Mesinger:2005ah,Bringmann:2011ut,Chluba:2012we}, the amplitude could be much larger, so that it is possible to form the PBHs when the large curvature perturbations reenter the horizon.
While the inflation is now regarded as a crucial ingredient in modern standard cosmology model, it is quite important to give constraints on curvature perturbations on small scales in order to identify a successful inflationary model, besides the constraints from CMB observations on large scales.
The observations of PBHs and stochastic gravitational wave background provide such a possibility in the future to give further constraints on inflationary models.

In this paper, we consider a single-field inflationary model with a very small periodic structure upon the potential.
Such a model can naturally arise from brane inflation and axion inflation~\cite{Freese:1990rb,Dimopoulos:2005ac,Bean:2008na,McAllister:2008hb,Flauger:2009ab}.
In such a model, PBHs can be all of dark matter, or a fraction of dark matter in our Universe, depending on the model parameters.
And the associated stochastic gravitational wave background could be detected by further GW detectors, such as the pulsar timing array and the space-based gravitational wave detectors.
For convenience, we set $c=\hbar=8\pi G= M_{\mathrm{Pl}} = 1$ throughout this paper.

\section{Parametric amplification of curvature perturbations}
We firstly revisit how the primordial power spectrum of curvature perturbations is generated in the standard single-field slow-roll inflation with a potential $V(\phi)$.
Usually the comoving curvature perturbation $\mathcal{R}$ is used to characterize primordial inhomogeneities, and it's convenient to introduce a variable $u=-z \mathcal{R}$, where $z=a\dot{\phi}/H$ and the overdot denotes a derivative with respect to the cosmic time $t$.
The equation of motion for a Fourier mode $u_{k}$ is given by the Mukhanov-Sasaki equation~\cite{Kodama:1985bj,Mukhanov:1990me}:
\begin{equation}
u''_{k}+(k^{2}-z''/z)u_{k}=0,
\end{equation}
where the prime denotes a derivative with respect to the conformal time $\tau$. The effective mass term, $z''/z$, can be written as~\cite{Stewart:1993bc}
\begin{equation}\label{eq:zppz}
     \frac{z''}{z}=2 a^{2} H^{2} \left( 1+\frac{3}{2}\delta +\epsilon +\frac{1}{2}\delta^{2} +\frac{1}{2}\epsilon\delta +\frac{1}{2H}\dot{\epsilon} +\frac{1}{2H}\dot{\delta} \right),
\end{equation}
where $\epsilon \equiv -\dot{H}/H^{2}$ and $\delta \equiv \ddot{\phi}/H\dot{\phi} $.
In the standard slow-roll inflation, $\epsilon$ and $\delta$ are less than $1$ and can be assumed as small constants during inflation, which gives $z''/z \approx 2a^{2}H^{2}$.
By setting the initial condition of $u_{k}(\tau)$ to be the Bunch-Davies vacuum, one can get the nearly scale-invariant power spectrum $\mathcal{P}_{\mathcal{R}}(k)=(H^{2}/2\pi\dot{\phi})^{2}|_{k=aH}$, which explains the measurements on CMB scales.

But in order to generate a non-negligible number of PBHs, the amplitude of curvature perturbations on small scales needs to be several orders of magnitude larger than the one on CMB scales~\cite{Young:2019yug,Sato-Polito:2019hws}.
This can be achieved through several ways, such as running mass inflationary models~\cite{Stewart:1996ey,Stewart:1997wg,Leach:2000ea}, multiple-field inflationary models~\cite{Cheng:2018yyr,Gao:2018pvq}, ultra-slow-roll inflationary models~\cite{Germani:2017bcs,Nakama:2018utx,Byrnes:2018txb} or sound speed resonance~\cite{Cai:2018tuh,Cai:2019jah,Chen:2019zza}.
In this paper, we propose a novel possibility in which curvature perturbations can be amplified by a small periodic structure in the inflationary potential.

\begin{figure}[htpb]
    \centering
    \includegraphics[width=0.6\textwidth]{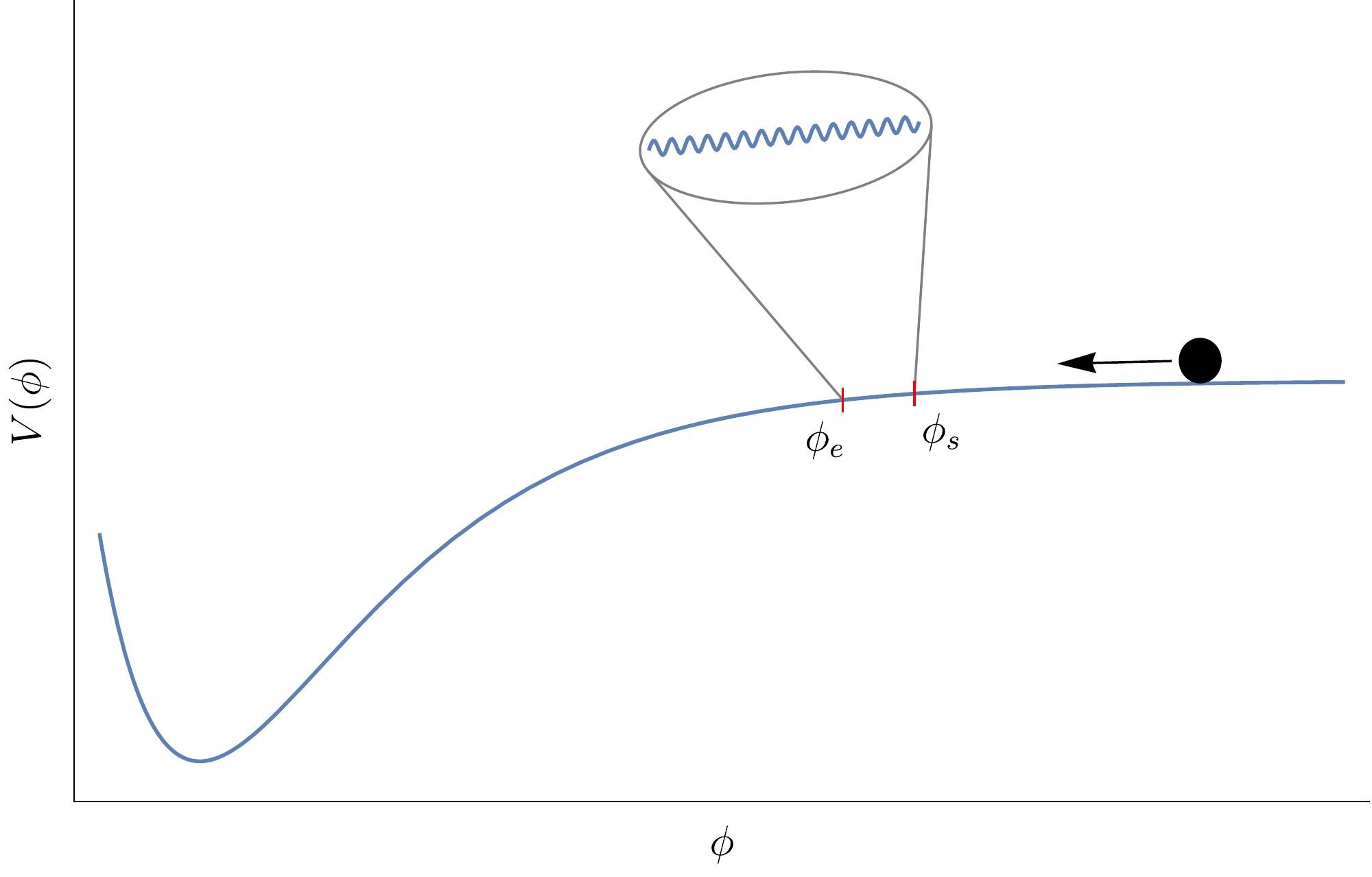}
    \caption[]{Schematic diagram of the periodic structure of single-field slow-roll inflationary potential.}
    \label{fig:res_strc}
\end{figure}

For the purpose of demonstration, let us consider a simple model.
Suppose there is a small part with a periodic structure in the inflationary potential as shown in Fig.\ref{fig:res_strc}, which can be written as
\begin{equation}
    \begin{aligned}
        V(\phi)&=\bar{V}(\phi)+\delta V(\phi)\\
               &=\bar{V}(\phi)+\xi \cos (\phi/\phi_{*}) \Theta(\phi;\phi_{s},\phi_{e}),
    \end{aligned}
\end{equation}
where $\bar{V}$ is the single-field slow-roll inflationary potential, $\delta V$ is a small contribution with a periodic structure to the potential, and $\delta V \ll \bar{V}$.
$\xi$ describes the magnitude of the structure, $\phi_{*}$ characterizes the period of the structure, $\phi_{s}$ and $\phi_{e}$ denote the starting and ending points of the structure respectively, and $\Theta(\phi;\phi_{s},\phi_{e})\equiv \theta(\phi_{e}-\phi)\theta(\phi-\phi_{s})$ where $\theta$ is the Heaviside step function.
This model can arise naturally in brane inflation~\cite{Bean:2008na} or axion monodromy inflation~\cite{McAllister:2008hb,Flauger:2009ab}.

We consider the case that the small periodic structure does not affect the background evolution, and focus on the period when the inflaton goes through from $\phi_{s}$ to $\phi_{e}$ (any quantity with subscript ${s}$ means it is evaluated when $\phi=\phi_{s}$ except specific definition, likewise for subscript ${e}$).
In this case, $\dot{\phi}$ and  $\bar{V}_{,\phi}$ evolve with time slowly while $\ddot{\phi}$ and  $\delta V_{,\phi}$ oscillate with time fastly, where the subscript denotes the derivative with respect to $\phi$.
Therefore the Klein-Gordon equation of the inflaton
\begin{equation}
    \ddot{\phi}+3H\dot{\phi}+V_{,\phi}=0,
\end{equation}
can be divided into
\begin{align}
3H\dot{\phi}+\bar{V}_{,\phi}\approx 0,\\
\ddot{\phi}+\delta V_{,\phi}\approx 0,
\end{align}
and this division will also be validated by the numerical simulation.
Thus
\begin{equation}
\phi \approx \phi_{s} + \dot{\phi}_{s}(t-t_{s})
\end{equation}
is given by the background potential $\bar{V}$, and the potential $\delta V$ gives
\begin{equation}
\ddot{\phi} \approx \xi \sin(\phi/\phi_{*})/\phi_{*}.
\end{equation}

Note that $H \ll 1$, because of this structure $\delta V$, the dominant term of the effective mass $z''/z$ in Eq.\eqref{eq:zppz} becomes the last one, which gives
\begin{equation}
    \frac{z''}{z} \approx a^{2} H \dot{\delta}.
\end{equation}
Recalling the definition of $\delta$, we have
\begin{equation}
    \delta \equiv \frac{\ddot{\phi}}{H\dot{\phi}} \approx \frac{1}{H\dot{\phi_{s}}} \frac{\xi}{\phi_{*}} \sin(\frac{\phi}{\phi_{*}}),
\end{equation}
and
\begin{equation}
    \dot{\delta} = \frac{d\delta}{d\phi}\dot{\phi} \approx \frac{1}{H}\frac{\xi}{\phi_{*}^{2}}\cos(\frac{\phi}{\phi_{*}}),
\end{equation}
which lead to
\begin{equation}
    \frac{z''}{z} \approx a^{2} \frac{\xi}{\phi_{*}^{2}} \cos (\frac{\phi}{\phi_{*}}).
\end{equation}
Therefore the equation of motion of $u_{k}$ can be written as
\begin{equation}
    \ddot{u}_{k} + H \dot{u}_{k} + \left[ \frac{k^{2}}{a^{2}} - \frac{\xi}{\phi_{*}^{2}}\cos(\frac{\phi}{\phi_{*}}) \right] u_{k} = 0.
\end{equation}

On the small scales ($k \gg a H$) which correspond to the scales with PBH formation, the $H\dot{u}_{k}$ term can be neglected, which gives
\begin{equation}\label{eq:Mathieu}
    \frac{d^{2}u_{k}}{dx^{2}}+ \left[ A_{k}(x)-2q\cos 2x \right] u_{k} =0,
\end{equation}
where $k_{*}=\frac{|\dot{\phi}_{s}|}{2\phi_{*}}$, $x=\frac{\dot{\phi}_{s}}{2\phi_{*}} t+\frac{\phi_{s}-\dot{\phi}_{s}t_{s}}{2\phi_{*}}$, $A_{k}(x)=\frac{k^{2}}{k_{*}^{2} a^{2}}$, $q=2\xi/\dot{\phi}_{s}^{2}$.\footnote{The inflaton has order Hubble quantum kicks in each e-fold, one may wonder  whether it will break the parameter resonance. Here we show that the 
quantum kicks from the frozen modes have negligible effects on Eq.~\eqref{eq:Mathieu}.  Note that $\dot{\phi}_{s}$ in Eq.(2.14) can be written as $\dot{\phi}_{s}=\dot{\bar{\phi}}_{s}+\delta\dot{\phi}_{s}$, where $\dot{\bar{\phi}}_{s}$ is from the $k=0$ mode and $\delta\dot{\phi}_{s}$ is from the $0<k/a<H$ modes. Namely,  $\dot{\bar{\phi}}_{s}$  is from the 
background evolution, while $\delta\dot{\phi}_{s}$ is from the quantum kick.  Since the quantum kicks satisfy $\delta\phi_{s}\sim H$, $\delta\dot{\phi}_{s}\sim H\delta\phi_{s}\sim H^{2}$ for the modes with $k\sim H$. For the parameter sets in Tab.~\ref{tab:para}, $\delta\dot{\phi}_{s}\sim H^{2}\sim 10^{-10}$ which is much smaller than $\dot{\bar{\phi}}_{s}\sim 10^{-7}$.
therefore,  the quantum kick effect $\delta\dot{\phi}_{s}$  can be neglected.}
This is the Mathieu equation except $A_{k}$ varies with $x$, where the parametric instability may present for some ranges of $k$.\footnote{This equation is also studied in the studies on preheating~\cite{Traschen:1990sw,Kofman:1997yn,Allahverdi:2010xz}.}
In some parameter space, we have  $0< q \ll 1$, and the resonance bands are located in narrow ranges around $A_{k} \sim n$.
Since the first band ($A_{k} \sim 1$) is the most enhanced one, in what follows we will focus on this band.

\begin{figure}[htpb]
    \centering
    \includegraphics[width=0.6\textwidth]{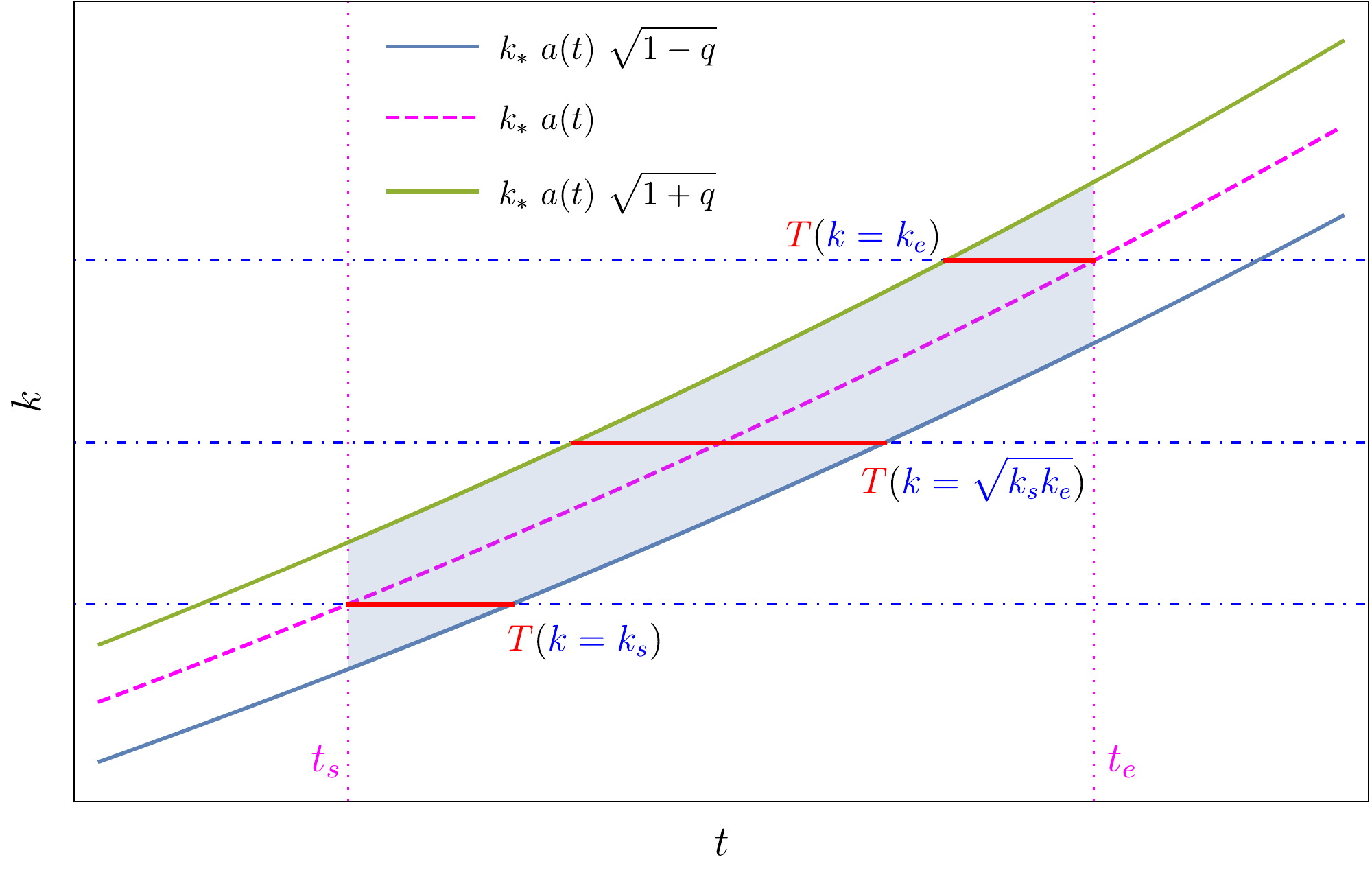}
    \caption[]{Schematic diagram for the first resonance band of Eq.\eqref{eq:Mathieu}(shaded region) and the total time that a $k$ mode can stay in the resonance band  $T(k)$ (red solid line). }
    \label{fig:Tk}
\end{figure}

For a given $k$ mode, it will stay in the resonance band only if
$1-q < A_{k} <1+q$.
A significant difference between Eq.\eqref{eq:Mathieu} and the Mathieu equation is that $A_{k}$ slowly varies with $x$, which means the total time that a given $k$ mode can stay in the resonance band is finite, and is given by
\begin{equation}\label{eq:Tk}
    T(k)=\left\{
    \begin{aligned}
        &0,&k \le k_{s}\sqrt{1-q},\\
        & \min(t_{e},t_{ek}) - \max(t_{s},t_{sk}),\quad & k_{s}\sqrt{1-q} < k < k_{e}\sqrt{1+q},\\
        &0,&k \ge k_{e}\sqrt{1+q},
    \end{aligned}
    \right.
\end{equation}
where $k_{s} \equiv k_{*}a_{s}$, $k_{e} \equiv k_{*}a_{e}$, and $t_{sk}$, $t_{ek}$ are the time that the mode enters and leaves the resonance band, which satisfy $A_{k}(t_{sk}) = 1+q$, $A_{k}(t_{ek}) = 1-q$. Eq.\eqref{eq:Tk} can be easily understood with the help of Fig.\ref{fig:Tk}.

In the resonance band, $u_{k}$ grows exponentially with rate $\mu_{k}$ which is given by
\begin{equation}
    \mu_{k}(t) =\Re \left(  \sqrt{\left( \frac{q}{2} \right)^{2}-\left( \frac{k}{k_{*}a(t)}-1 \right)^{2}}  \right),
\end{equation}
where $\Re(y)$ denotes the real part of $y$.
Therefore during the period from $t_{s}$ to $t_{e}$, $u_{k}$ can be amplified by
\begin{equation}\label{eq:Ak}
    \mathcal{A}(k) \equiv |u_{k}(t_{e})| / |u_{k}(t_{s})| \approx \exp \left( \int_{t_{s}}^{t_{e}} \mu_{k}(t) k_{*} dt \right),
\end{equation}
and the power spectrum of curvature perturbations can be written as
\begin{equation}
    \mathcal{P}_{\mathcal{\mathcal{R}}}(k) \approx  A_{s} \left( \frac{k}{k_{p}} \right)^{n_{s}-1} \mathcal{A}^{2}(k),
\end{equation}
where $A_{s}$ is the amplitude of the power spectrum as in the standard inflation, and $n_{s}$ is the scalar spectral index at the pivot scale $k_{p} \simeq 0.05~\Mpc^{-1}$.

To study further, we consider the Starobinsky potential
\begin{equation}
    \bar{V}(\phi) = \Lambda^{4} \left[ 1-\exp(-\sqrt{2/3}\phi) \right]^{2},
\end{equation}
where $\Lambda=0.0032$ is set to match the CMB experiments and we set the e-folding number when the pivot scale leaves horizon as $N_{\mathrm{CMB}}=60$~\cite{Akrami:2018odb}.

\begin{figure}[htpb]
    \centering
    \includegraphics[width=0.32\textwidth]{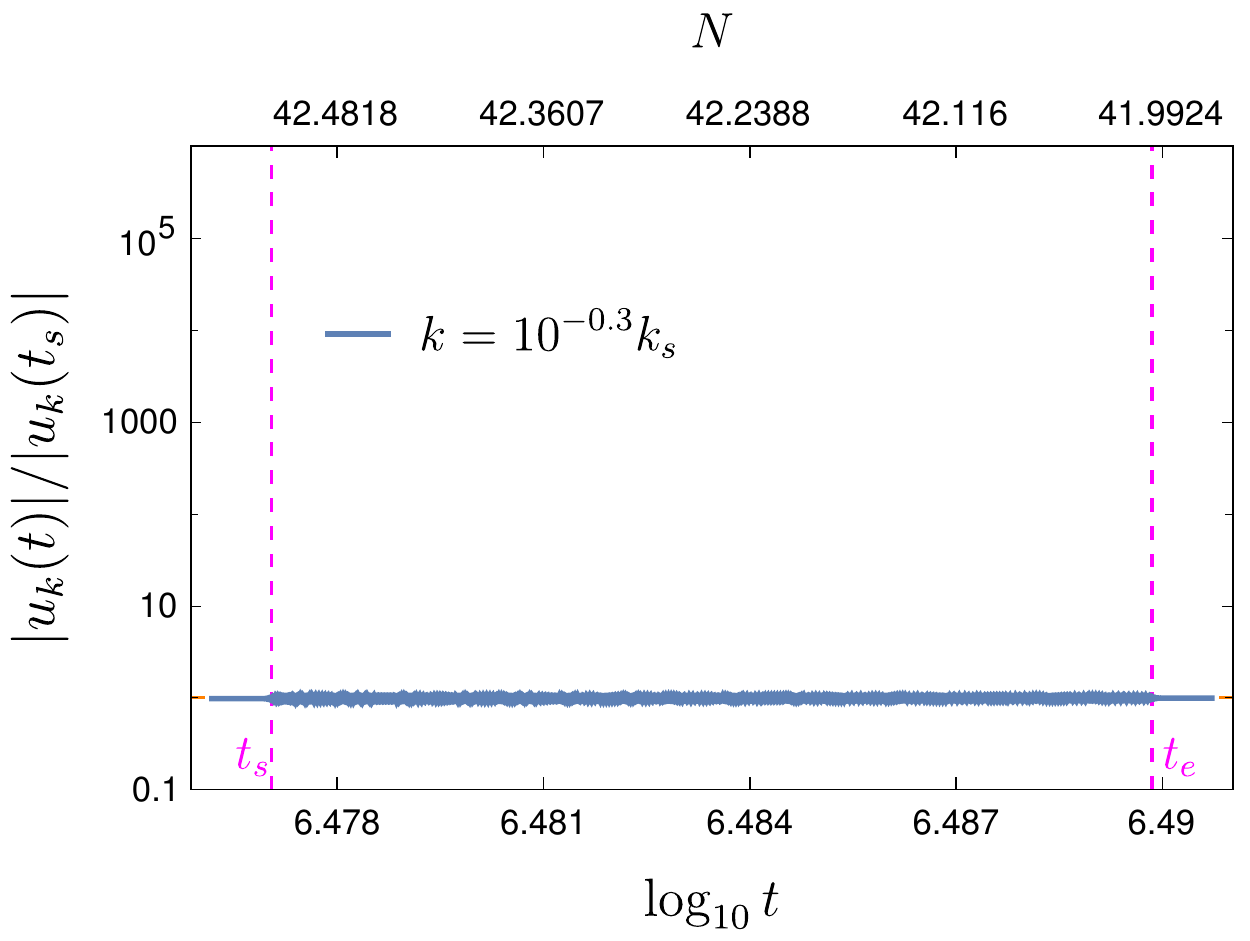}
    \includegraphics[width=0.32\textwidth]{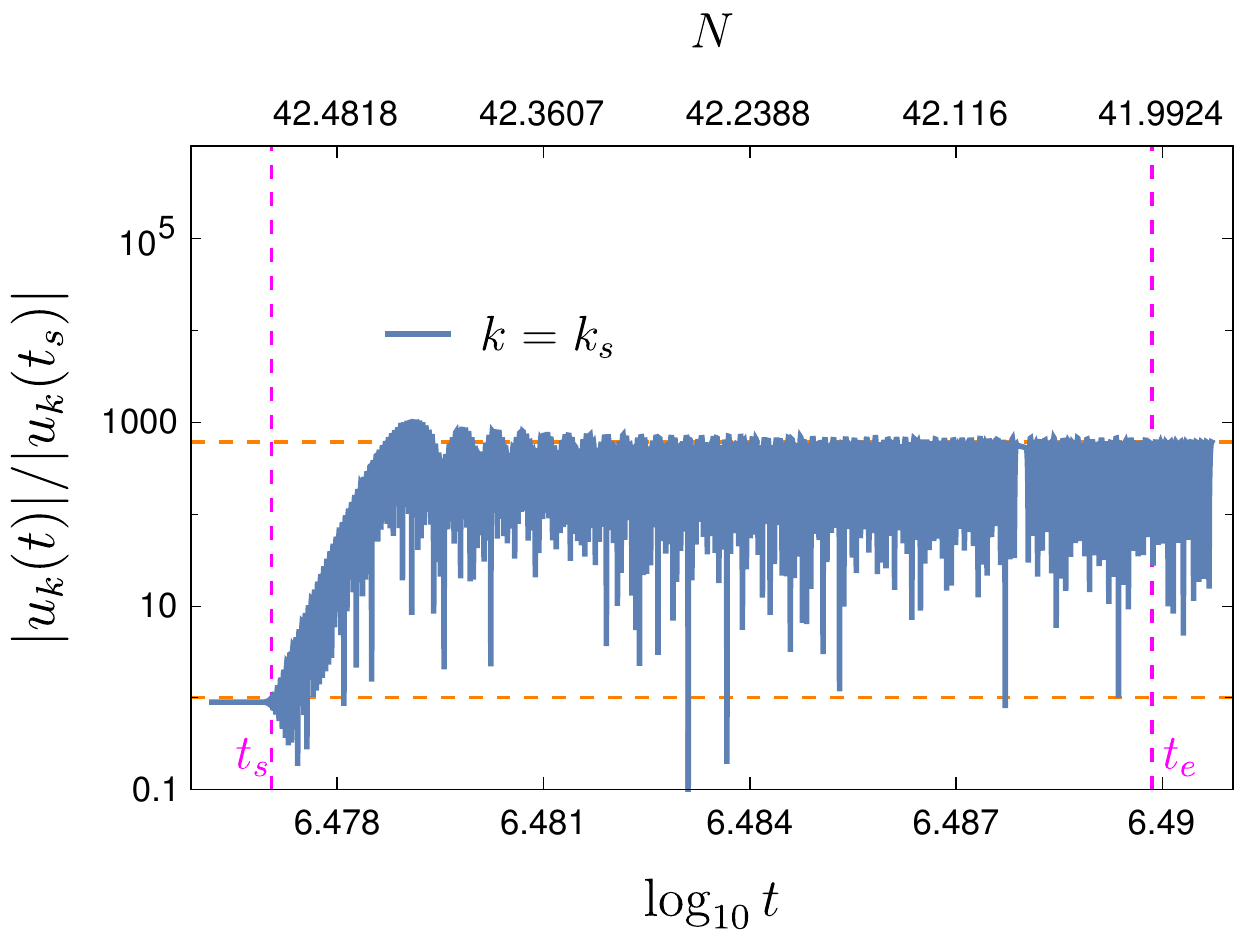}
    \includegraphics[width=0.32\textwidth]{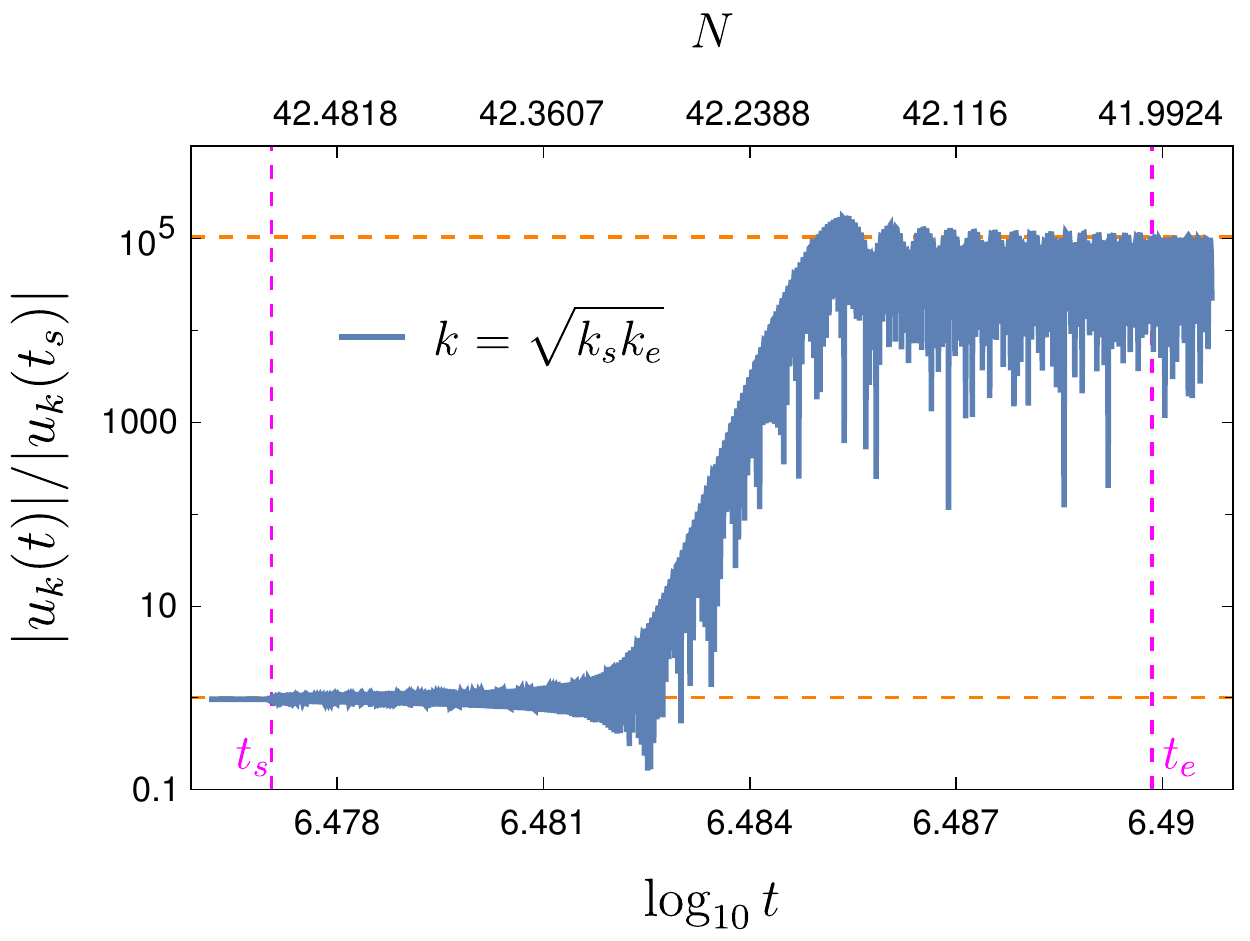}\\
    \includegraphics[width=0.32\textwidth]{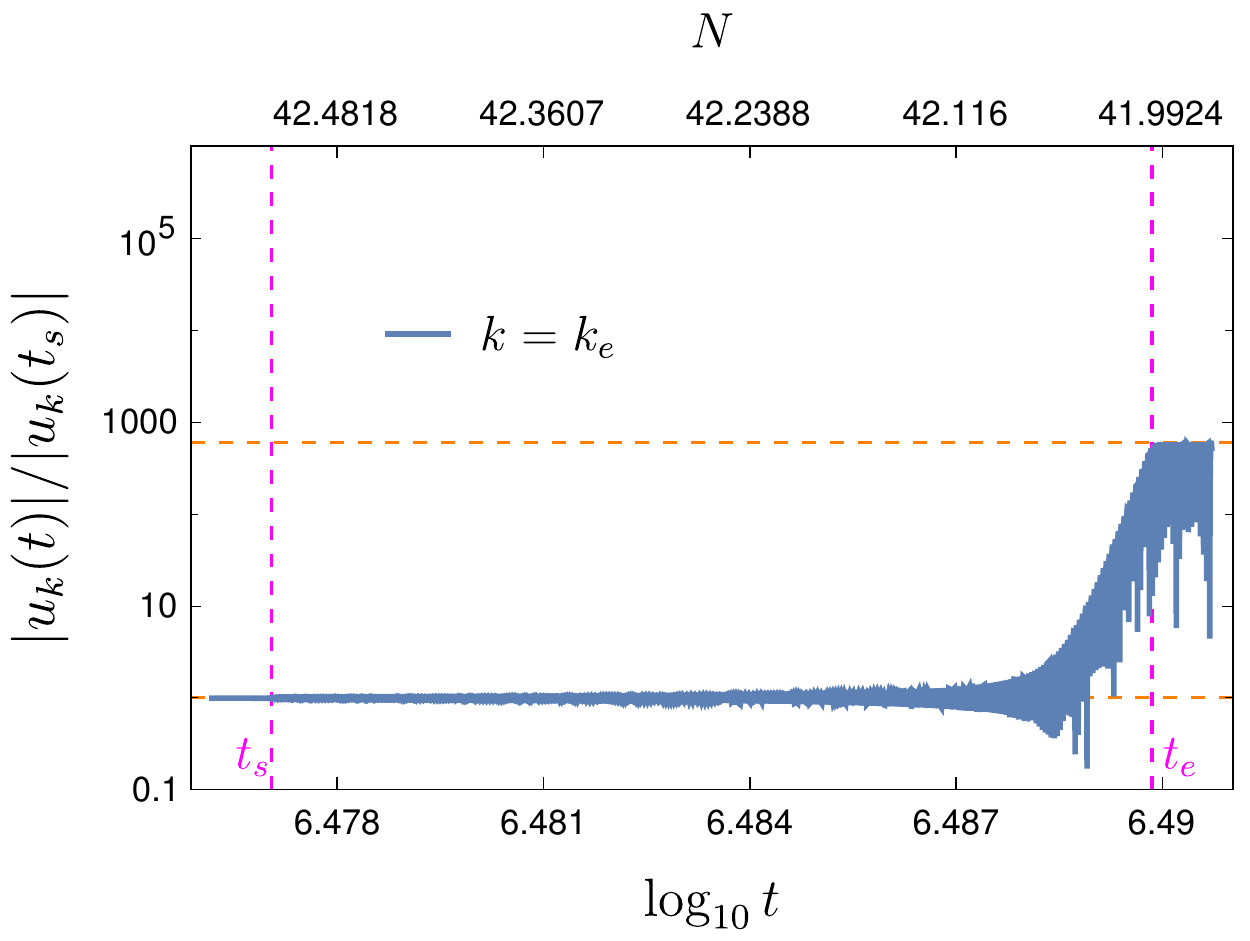}
    \includegraphics[width=0.32\textwidth]{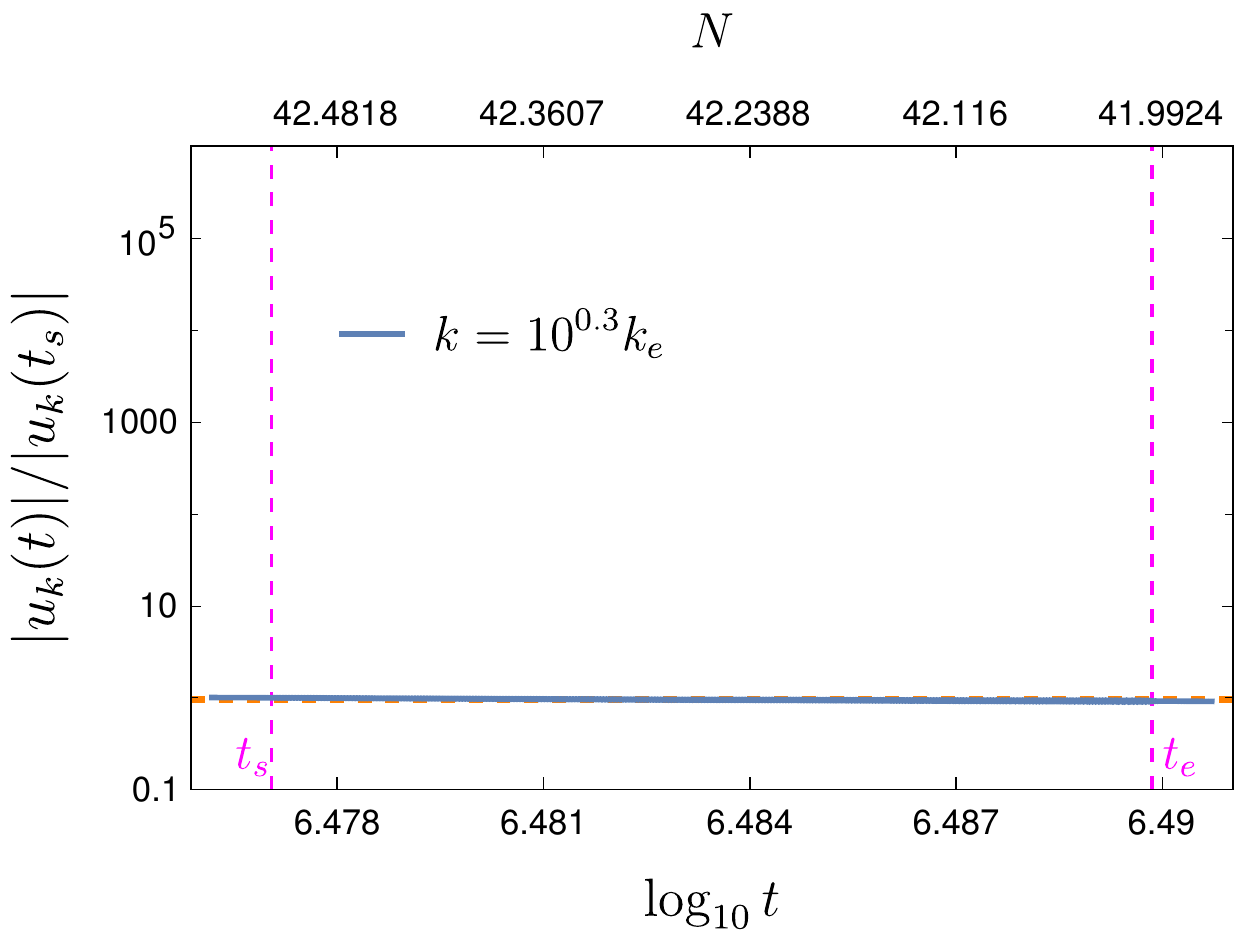}
    \includegraphics[width=0.32\textwidth]{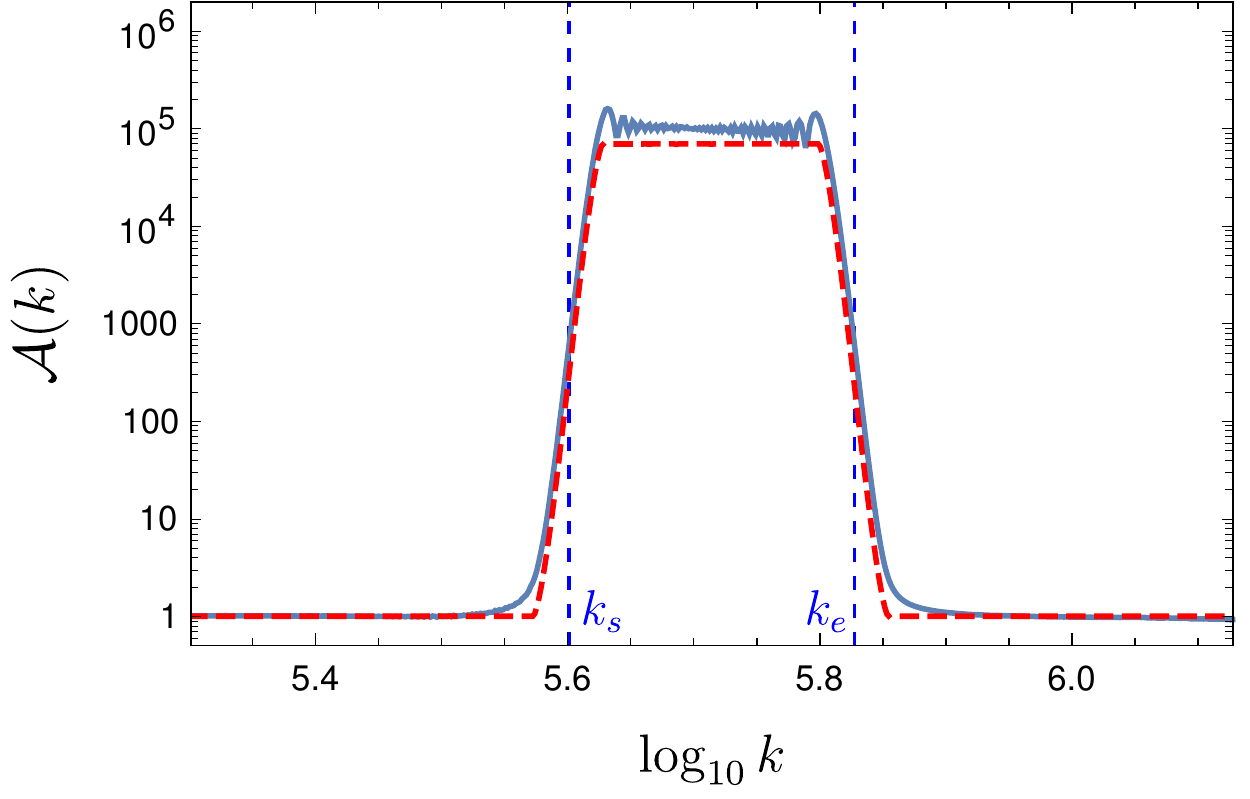}
    \caption[]{Numerical solutions of Mukhanov-Sasaki equation with the parameter set 1 in Tab.\ref{tab:para}.
        The evolution of $|u_{k}(t)| / |u_{k}(t_{s})|$ with the cosmic time $t$ for different  $k$ modes: $k=10^{-0.3}k_{s}$(top left), $k=k_{s}$(top middle), $k=\sqrt{k_{s} k_{e}}$(top right), $k=k_{e}$(bottom left) and $k=10^{0.3}k_{e}$(bottom middle).
The amplified ratio $\mathcal{A}(k)$ of the mode $|u_{k}|$ crossing the resonance band is shown in the bottom right panel, where the blue solid line represents the numerical result while the red dashed line represents the analytical approximate result Eq.\eqref{eq:Ak}.
    }
    \label{fig:uk}
\end{figure}

We solve the Mukhanov-Sasaki equation numerically and the results are shown in Fig.\ref{fig:uk}.
As we can see, the mode with $k \le k_{s}\sqrt{1-q}$ or $k \ge k_{e}\sqrt{1+q}$, such as $k=10^{-0.3}k_{s}$ or $k=10^{0.3}k_{e}$, is not amplified because $T(k)=0$ for this mode, but the mode with $k_{s}\sqrt{1-q} < k < k_{e}\sqrt{1+q}$, such as $k=k_{s}$, $k=k_{e}$, $k=\sqrt{k_{s} k_{e}}$, can be exponentially amplified because $T(k) \not= 0$.
The amplified ratio $\mathcal{A}(k)$ of the mode $|u_{k}|$ crossing the resonance band is shown in the bottom right panel of Fig.\ref{fig:uk}, where the blue solid line represents the numerical result while the red dashed line represents the analytical approximate result Eq.\eqref{eq:Ak}.
It is worth noticing that there is a plateau in $\mathcal{A}(k)$, this is because if $k_{s}\sqrt{1+q} < k_{e}\sqrt{1-q}$, then $T(k)=(2H)^{-1} \ln \frac{1+q}{1-q}$ of these modes with $k_{s}\sqrt{1+q} < k < k_{e}\sqrt{1-q}$ is independent of $k$.
Another interesting thing is that there are some ripples in the plateau, which originate from the phase of $u_{k}$ when it leaves the resonance band, and may significantly affect the abundance of PBHs, since the mass function of PBHs is extremely sensitive to $\mathcal{P}_{\mathcal{R}}$ as we will see in the next section.
We leave the detailed discussions to the future work.

\begin{table}[htpb]
    \centering
	\begin{tabular}{|c|c|c|c|c|}
		\hline
		Set&$\xi$& $\phi_{*}$ &$\phi_{s}$ &$\phi_{e}$ \\
		\hline
		$1$ &$1.70\times 10^{-15}$ & $8.00\times 10^{-6}$ &$4.9878$  &$4.9731$\\
		\hline
		$2$ &$1.23 \times 10^{-15}$ & $6.64 \times 10^{-6}$ &$5.2118$  &$5.2088$\\
		\hline
		$3$ &$1.38 \times 10^{-15}$ & $7.39 \times 10^{-6}$ &$5.0761$  &$4.8782$\\
		\hline
		$4$ &$2.29 \times 10^{-15}$ & $9.40 \times 10^{-6}$ &$4.7920$  &$4.7880$\\
		\hline
	\end{tabular}
	\caption{The parameter sets we choose as examples. }
	\label{tab:para}
\end{table}

\begin{figure}[htpb]
    \centering
    \includegraphics[width=0.6\textwidth]{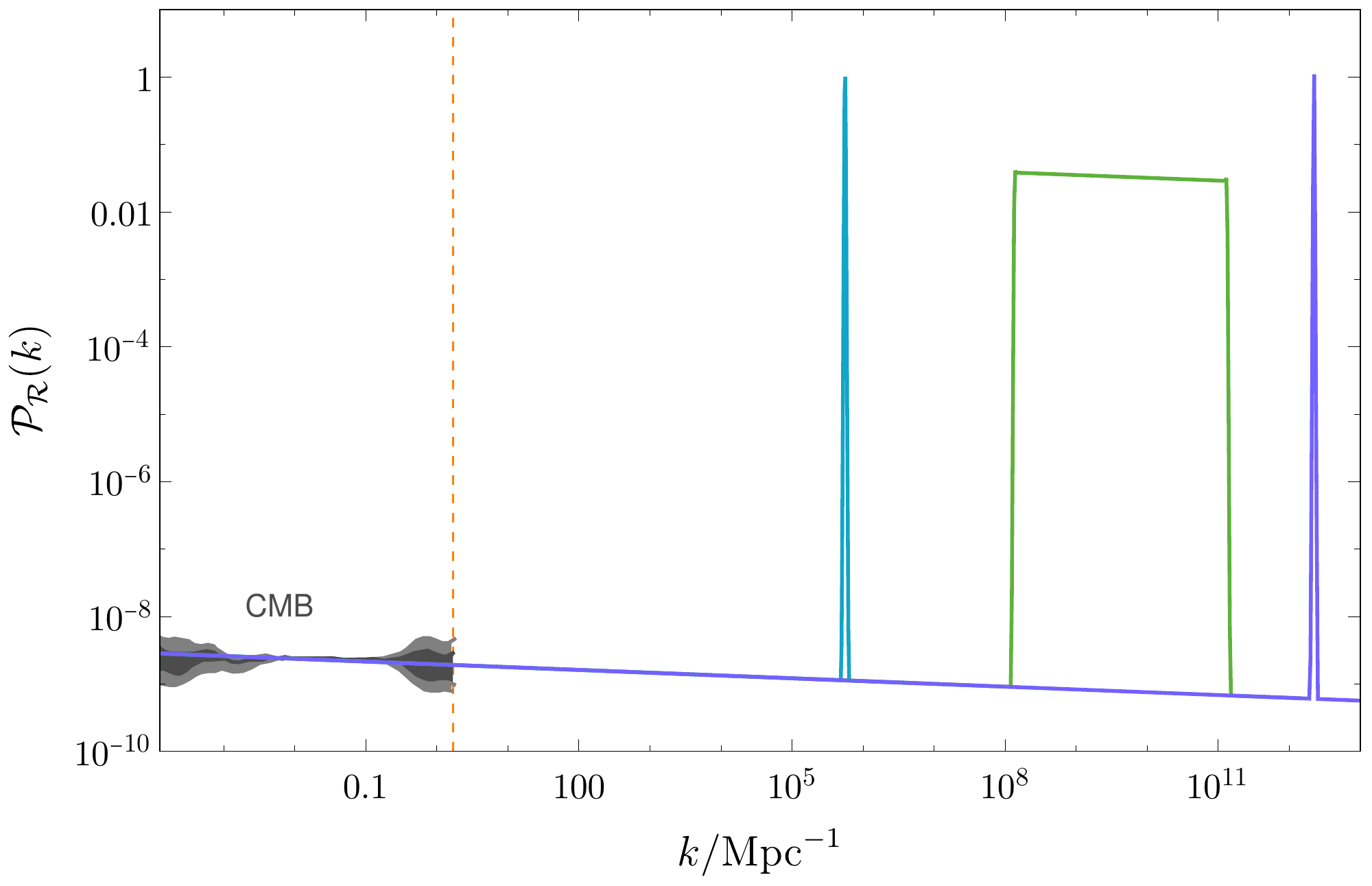}
    \caption[]{The power spectrum of curvature perturbations originates from the periodic structure of the inflationary potential with the parameter set 2(blue), set 3(green), set 4(purple) in Tab.\ref{tab:para}, and constraint from cosmic microwave background (CMB~\cite{Bringmann:2011ut}).}
    \label{fig:PR}
\end{figure}

For simplicity, we use the analytical approximate formula of $\mathcal{A}(k)$, i.e. Eq.\eqref{eq:Ak}, to evaluate the power spectrum of curvature perturbations, and show the results of three sets of parameters $(\xi, \phi_{*}, \phi_{s}, \phi_{e})$ in Fig.\ref{fig:PR}.

\section{PBH formation}
Now we study the formation of PBHs due to the enhancement of the power spectrum of curvature perturbations on small scales.
After leaving the horizon during inflation, curvature perturbations get frozen until reentry.
The overdense regions with the average density larger than a critical value could collapse gravitationally into black holes.
Usually the abundance of PBHs is described by the PBH mass fraction of horizon mass $m_{\mathrm{H}}$ at the formation time, which can be written as an integration of the Gaussian distribution of perturbations~\footnote{Non-Gaussianity is generated in this kind of models~\cite{Chen:2008wn,Flauger:2010ja,Hannestad:2009yx}. In Refs.~\cite{Young:2019yug,DeLuca:2019qsy}, it is found that non-Gaussianity can affect the PBH abundance. However, using the method of~\cite{Franciolini:2018vbk}, we confirmed the non-Gaussianity does not affect the mass function for the parameters we used.}
\begin{equation}\label{eq:beta}
    \beta(M) \equiv \frac{\rho_{\mathrm{PBH}}(M)}{\rho_{\mathrm{tot}}}=\gamma \ \erfc \left( \frac{\Delta_{c}}{\sqrt{2 \sigma_{M}^{2}}} \right),
\end{equation}
where $\gamma \simeq 0.2$ is a correction factor~\cite{Carr:1975qj}, $\Delta_{c} \simeq 0.45$ is the critical density contrast above which PBHs will form~\cite{Musco:2004ak},~\footnote{This is usually used as a representative value. In fact, it depends on the shape of the density profile \cite{Musco:2018rwt}.}  and $ \sigma_{M}^{2} =  \int_{0}^{\infty} (dq/q) \tilde{W}^{2}(R=k_{M}^{-1}, q)  \frac{16}{81}  (q/k_{M})^{4} T^{2}(\eta=k_{M}^{-1}, q) \mathcal{P}_{\mathcal{R}}(q) $ is the variance of density contrast, where $\tilde{W}(R, q)=\exp(-q^{2}R^{2}/2)$ is the Fourier transform of a volume-normalized Gaussian window smoothing function, and $T(\eta, q)=3(\sin y - y \cos y)/y^{3}$ with $y \equiv q\eta/ \sqrt{3}$ is the transfer function.

\begin{figure}[htpb]
    \centering
    \includegraphics[width=0.6\textwidth]{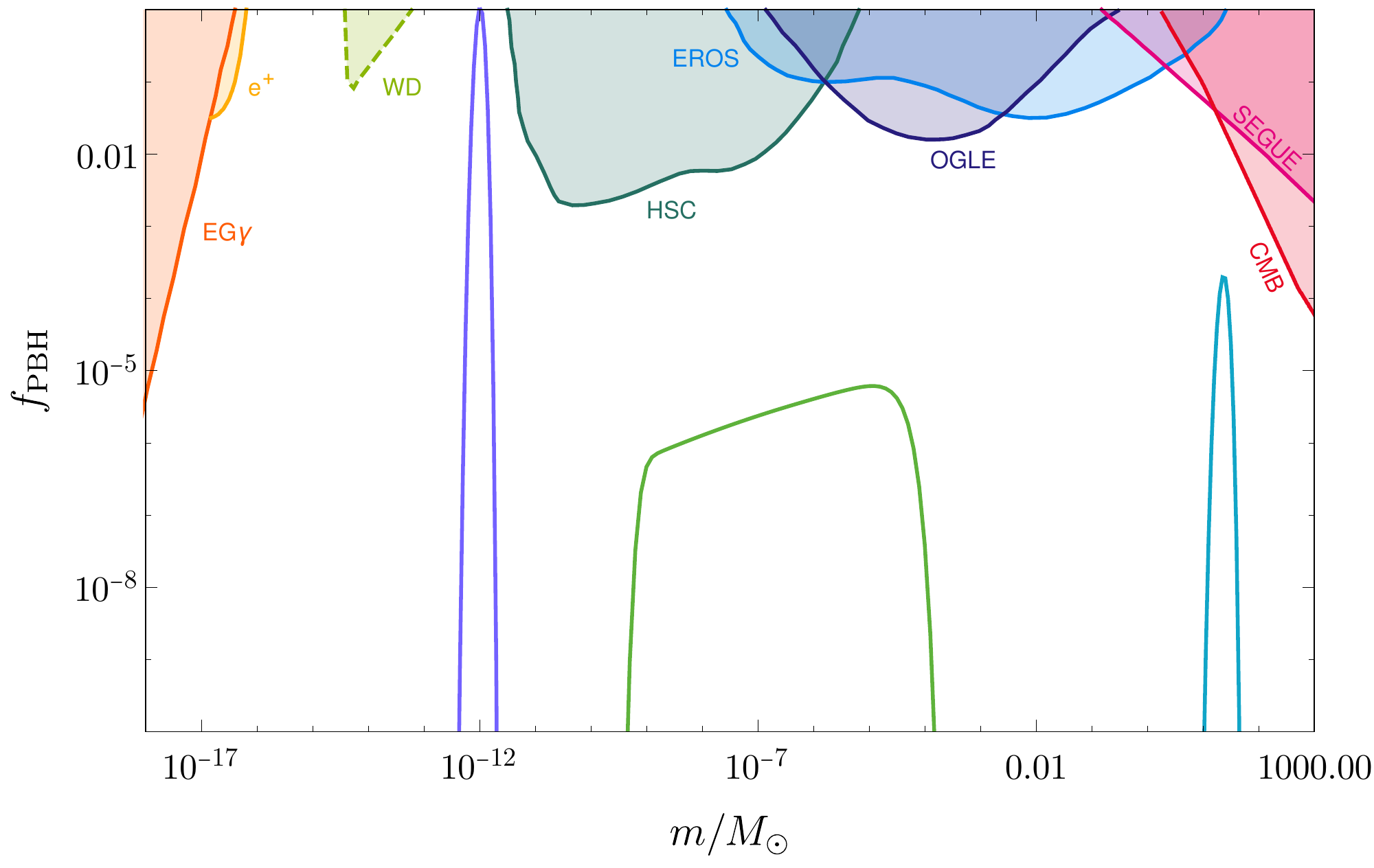}
    \caption[]{The fraction of PBHs in DM $f_{\mathrm{PBH}}$, produced by the periodic structure of the inflationary potential with the parameter set 2(blue), set 3(green), set 4(purple) in Tab.\ref{tab:para}.
        Here the constraints are from the extragalactic gamma ray background (EG$\gamma$ \cite{Carr:2009jm}), galactic 511 keV line ($e^{+}$ \cite{DeRocco:2019fjq,Laha:2019ssq,Dasgupta:2019cae}), white dwarf explosions (WD \cite{Graham:2015apa}, see \cite{Montero-Camacho:2019jte} for criticism), gravitational lensing events (HSC \cite{Niikura:2017zjd}, EROS \cite{Tisserand:2006zx}, OGLE \cite{Niikura:2019kqi}), dynamical effects (SEGUE \cite{Koushiappas:2017chw}), and cosmic microwave background (CMB~\cite{Poulin:2017bwe}).
    }
    \label{fig:fPBH}
\end{figure}

One can define the fraction of PBHs in the DM at present as~\cite{Sasaki:2018dmp}
\begin{equation}
        f_{\mathrm{PBH}}(M) \equiv \frac{\Omega_{\mathrm{PBH}}(M)}{\Omega_{\mathrm{DM}}} = 2.7 \times 10^{8} \left( \frac{\gamma}{0.2} \right)^{1/2} \left( \frac{g_{*,\mathrm{form}}}{10.75} \right)^{-1/4} \left( \frac{M}{M_{\odot}} \right)^{-1/2} \beta(M) ,
\end{equation}
where $g_{*,\mathrm{form}}$ is the relativistic degree of freedom when PBHs form.
It is found that the parameter space of the periodic structure of the inflationary potential is quite wide in this model.
We calculate $f_{\mathrm{PBH}}$ for three sets of parameters listed in Tab.\ref{tab:para} and show the results in Fig.\ref{fig:fPBH}.
For the parameter set 4, PBHs with mass centered at $10^{-12}$ solar mass may constitute all dark matter.
PBHs from the parameter set 2 correspond to the case with the assumption that all black holes detected by LIGO/VIRGO are PBHs~\cite{Cai:2019elf}.\footnote{
In fact, the calculations of PBH above are rough approximations.
To get more accurate results, one needs to consider more details of PBH formation, such as the critical collapse~\cite{Niemeyer:1997mt,Yokoyama:1998xd,Niemeyer:1999ak,Musco:2008hv,Musco:2011xj,Musco:2012au}, and the nonlinearity of the transfer function~\cite{Young:2019yug}.
However, since the mass function of PBHs mainly depends on the amplitude of $\mathcal{P}_{\mathcal{R}}$, we neglected the detailed calculations for demonstration.
}

\section{Induced GWs}
After reentering the horizon in the radiation-dominated epoch, scalar perturbations will inevitably lead to GWs at second order since they are coupled at nonlinear order~\cite{Ananda:2006af,Baumann:2007zm}.
Thus the induced GWs due to the enhancement of power spectrum of curvature perturbations may be detected if the enhancement is large enough.

In the conformal Newton gauge, neglecting vector perturbations and the anisotropic stress, the perturbed metric reads,
\begin{equation}
    ds ^{2} =a ^{2} ( \tau ) \left\{ - ( 1+2 \Phi ) d\tau ^{2} + \left[ ( 1-2\Phi ) \delta _{ij} + \frac{1}{2} h _{ij} \right] dx ^{i} dx ^{j} \right\},
\end{equation}
where $ \Phi $ is the Newton potential, and $ h _{ij} $ is the tensor perturbation.
Expanding $h_{ij}(\tau,\bm{x})$ as follows,
\begin{equation}
h_{ij}(\tau,\bm{x})=\int\frac{d^3k}{(2\pi)^{3/2}}\sum_{\lambda=+,\times}e_{ij}^\lambda(\hat k)h_{\bm{k},\lambda}(\tau)e^{i\bm{k}\cdot\bm{x}},
\end{equation}
where $+$ and $\times$ are two polarization modes, one can derive the equation of motion for $h_{ij}$ from the perturbed Einstein equations up to second order, and write it in the momentum space as
\begin{equation}\label{eq:eof}
    h_{\bm{k}} ''+2 \mathcal{H} h_{\bm{k}} '+k^{2} h_{\bm{k}} =2\mathcal{P}_{ij}^{lm}e^{ij}T_{lm}(\bm{k}, \tau ) ,
\end{equation}
for each polarization, where $\mathcal{H}$ is the conformal Hubble parameter, $\mathcal{P}_{ij}^{lm}$ is the projection operator to the transverse-traceless part, and the source term $T_{lm}( \bm{k} ,\tau )$ is of second order in scalar perturbations,
\begin{equation}
T_{lm}=-2\Phi\partial_l\partial_m\Phi+\partial_l\left(\Phi+\mathcal{H}^{-1}\Phi'\right)\partial_m\left(\Phi+\mathcal{H}^{-1}\Phi'\right).
\end{equation}
This equation can be solved with Green's function method and one can calculate the power spectrum of this tensor mode $\mathcal{P}_{h}$ from the two-point correlator of $h_{\bm{k}}$,
\begin{equation}
    \langle h_{\bm{k}}(\tau) h_{\bm{l}}(\tau)\rangle=\frac{2 \pi^{2}}{k^{3}} \delta^{(3)}(\bm{k}+\bm{l}) \mathcal{P}_{h}(\tau, k).
\end{equation}

Usually the stochastic GW is described by its energy density per logarithmic frequency interval normalized by the critical density,
\begin{equation}
    \Omega _{\mathrm{GW}} ( \tau ,k ) = \frac{1}{24} \left( \frac{k}{\mathcal{H ( \tau )}} \right) ^{2} \overline{\mathcal{P} _{h} ( \tau, k )},
\end{equation}
where the two polarization modes of GWs have been summed over, and the overline denotes average over several wavelengths.
Using the relation between curvature perturbations $\mathcal{R}$ and scalar perturbations $\Phi$ in the radiation-dominated era, $\Phi=-(2/3)\mathcal{R}$, one can calculate the spectrum of induced GWs until the matter-radiation equality.
After taking the thermal history of the Universe into consideration, one can get the GW spectrum at present by
\begin{equation}
        \Omega_{\mathrm{GW}}(k,\tau_{0}) = 2 \Omega_{r,0} \left( \frac{g_{*s}(T_{\mathrm{eq}})}{g_{*s}(T_{0})} \right)^{-\frac{4}{3}} \frac{g_{*r}(T_{\mathrm{eq}})}{g_{*r}(T_{0})} \Omega_{\mathrm{GW}}(k,\tau_{\mathrm{eq}}),
\end{equation}
where $g_{*s}$ and $g_{*r}$ are the effective degrees of freedom for entropy density and relativistic particles respectively, and we take $\Omega_{r,0} h^{2} = 4.2\times 10^{-5}$, $g_{*s}(T_{\mathrm{eq}})=3.91$, $g_{*r}(T_{\mathrm{eq}})=3.38$, $g_{*s}(T_{0})=3.91$, $g_{*r}(T_{0})=3.36$~\cite{Aghanim:2018eyx,Dodelson:2003ft,Kolb:1990vq}.

\begin{figure}[htpb]
    \centering
    \includegraphics[width=0.6\textwidth]{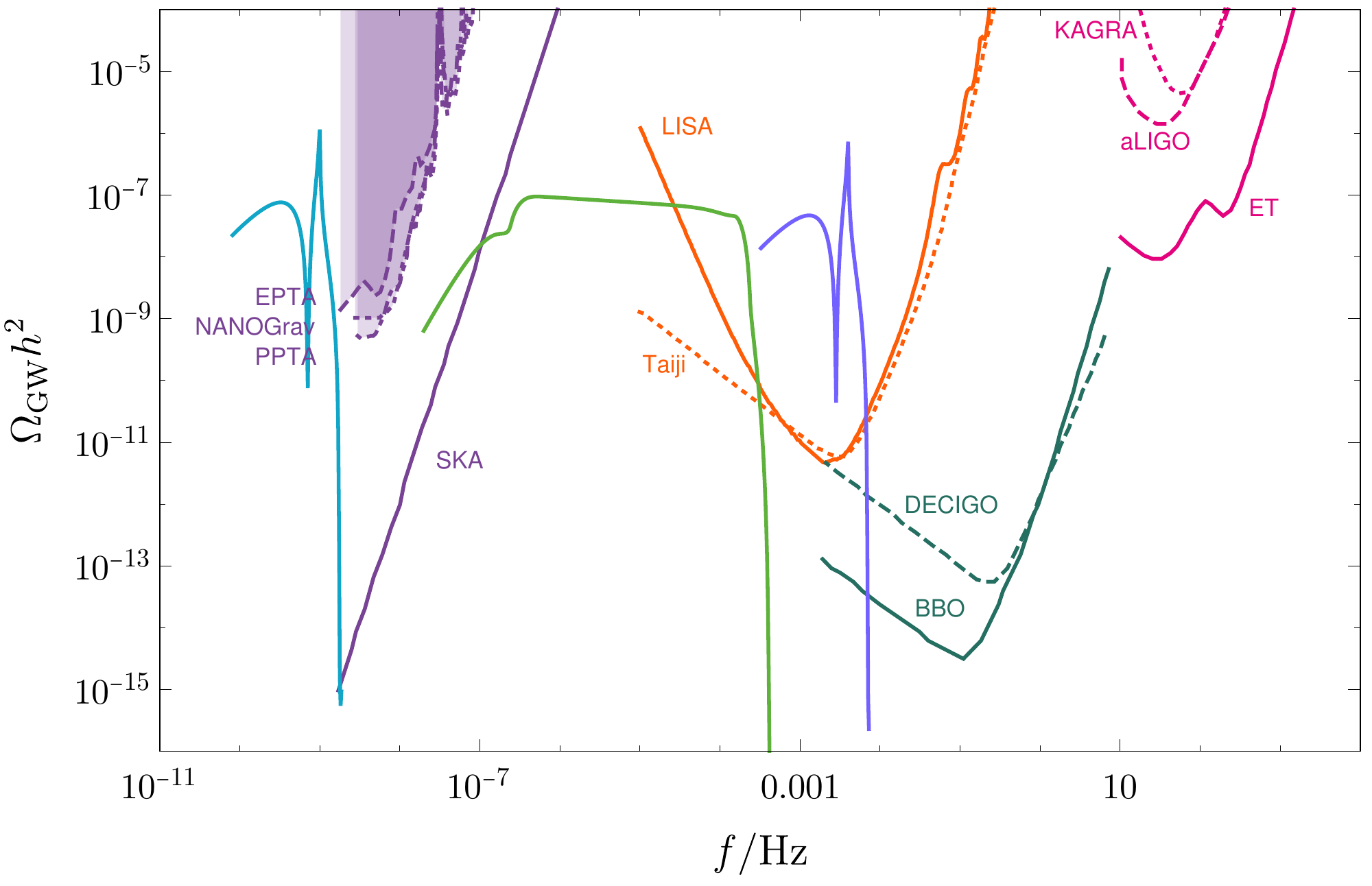}
    \caption[]{The energy spectrum of induced GWs with the parameter set 2(blue), set 3(green), set 4(purple) in Tab.\ref{tab:para}. We compare them with the constrain or the sensitivity curves of GW detectors including EPTA~\cite{Lentati:2015qwp}, NANOGrav~\cite{Arzoumanian:2018saf}, PPTA~\cite{Shannon:2015ect}, SKA~\cite{Carilli:2004nx}, LISA~\cite{Audley:2017drz}, Taiji~\cite{Guo:2018npi}, DECIGO~\cite{Kawamura:2011zz}, BBO~\cite{phinney2004big}, KAGRA~\cite{Somiya:2011np}, aLIGO~\cite{TheLIGOScientific:2014jea}, ET~\cite{Punturo:2010zz}. }
    \label{fig:fig-OmegaGWh2-pdf}
\end{figure}

In Fig.\ref{fig:fig-OmegaGWh2-pdf}, we plot the energy spectrum of GWs induced from curvature perturbations with the parameter sets in Tab.\ref{tab:para} .
Since the frequency of GWs depends on the comoving length scale at reentry, the induced GWs can be detected in the corresponding band of frequency,
then all the space-based and ground-based GW detectors have the possibility to detect $\Omega_{\mathrm{GW}}$.
We find that the narrow peak of $\mathcal{P}_{\mathcal{R}}$ in Fig.\ref{fig:PR} results in a sharp peak of $\Omega_{\mathrm{GW}}$, while the wide plateau of $\mathcal{P}_{\mathcal{R}}$ results in a plateau in $\Omega_{\mathrm{GW}}$.
In all those cases, $\Omega_{\mathrm{GW}}$ has a power-law behavior in the infrared region, as stated in \cite{Cai:2019cdl}, while $\Omega_{\mathrm{GW}}$ cuts off sharply in the ultraviolet region.
From the location of the peak and the edges of the plateau, one can inversely determine $\phi_{s}$ and $\phi_{e}$ of the inflationary potential.

\section{Conclusions}
In this paper we investigated the parametric amplification of curvature perturbations during inflation with a small periodic structure upon the potential, such a structure is common in axion and brane inflationary scenarios.
We found abundant phenomena which can be detected by future PBH and GW obsevations.
The PBH mass and GW frequency depend on the comoving length scale at reentry, while the width of the PBH mass function and plateau of $\Omega_{\mathrm{GW}}$ depend on the duration of the resonant period.
By choosing different parameters, the PBHs with a sharp mass function can constitute all dark matter, or are a fraction of dark matter, like the case with the assumption that black holes detected by LIGO/VIRGO are primordial ones, and the energy spectrum of induced GWs can have a characteristic sharp peak or plateau.

Recalling that the power spectrum of curvature perturbations in Fig.\ref{fig:PR} is obtained from analytical approximation Eq.\eqref{eq:Ak}, the accurate power spectrum should be obtained from numerical calculations and has some ripples corresponding to the bottom right panel of Fig.\ref{fig:uk}, and there are peaks near $k_{s}$ and $k_{e}$.
Since the mass function of PBHs is extremely sensitive to $\mathcal{P}_{\mathcal{R}}$, the mass function can have a similar multi-peak profile, which can be distinguished by future PBH detections.
Furthermore, using this mechanism we can give constraints to the small periodic structure upon the inflationary potential.
We leave these topics to further investigations.

\section*{Acknowledgments}
We thank Chengjie Fu, Shi Pi and Misao Sasaki for useful discussions.
This work is supported in part by the National Natural Science Foundation of China Grants Nos.11647601, 11690021, 11690022, 11821505, 11851302, 11947302, 11991052 and by the Strategic Priority Research Program of CAS Grant NO. XDB23010500 and No.XDB23030100, and by the Key Research Program of Frontier Sciences of CAS.

\bibliographystyle{JHEP}
\bibliography{003_ref}

\end{document}